\documentclass[12pt]{JHEP3}

\usepackage{epsf}
\usepackage{latexsym}
\usepackage{epsfig,multicol}

\def\NPB#1#2#3{Nucl. Phys. B{#1} (19#2) #3}

\def\yzero{\smash{\hbox{$y\kern-4pt\raise1pt\hbox{${}^\circ$}$}}}

\def\beq{\begin{equation}}
\def\eeq{\end{equation}}
\def\beqa{\begin{eqnarray}}
\def\eeqa{\end{eqnarray}}

\def\-{\hphantom{-}}
\def\ov{\overline}
\def\s2{\frac{1}{\sqrt2}}

\def\beq{\begin{equation}}
\def\eeq{\end{equation}}
\def\beqa{\begin{eqnarray}}
\def\eeqa{\end{eqnarray}}
\def\tr{{\rm tr \,}}

\def\IF{\relax{\rm I\kern-.18em F}}
\def\II{\relax{\rm I\kern-.18em I}}
\def\IP{\relax{\rm I\kern-.18em P}}
\def\IC{\relax\hbox{\kern.25em$\inbar\kern-.3em{\rm C}$}}
\def\IR{\relax{\rm \bf R}}
\def\IT{\relax{\rm \bf T}}
\def\IZ{\relax{\rm \bf Z}}

\def\Dsl{\,\raise.15ex\hbox{/}\mkern-13.5mu D} 

\newbox\pippobox

\title{Gauging Away the Strong CP  Problem}

\author{G. Aldazabal $^{1,2}$, L.~E.~Ib\'a\~nez $^{1}$ and  A. M. Uranga  
$^{1}$ \\
 	 $^1$ Departamento de F\'{\i}sica Te\'orica C-XI
	and Instituto de F\'{\i}sica Te\'orica  C-XVI,\\
	Universidad Aut\'onoma de Madrid,
	Cantoblanco, 28049 Madrid, Spain.\\
$^2$  Instituto Balseiro and Centro At\'omico
Bariloche, \\
8400 S.C. de Bariloche, (CNEA  and CONICET), Argentina
}

\preprint{\hepth{02mmddd}}
\preprint{FTUAM-02/15, IFT-UAM/CSIC-02-16,CAB-IB/ 2903002}

\abstract{
We propose a new solution to the strong-CP problem. It involves the 
existence of an unbroken gauged $U(1)_X$ symmetry whose gauge boson gets a
Stuckelberg mass term by combining with a pseudoscalar  field $\eta (x)$.
The latter has axion-like couplings to $F_{QCD}\wedge F_{QCD}$.
This system leads to  mixed gauge anomalies and we argue that they are
cancelled by the addition of an appropriate Wess-Zumino term, so that no
SM fermions need to be charged under $U(1)_X$. In this setup the axion and
$\theta$ parameter can be rotated away using the symmetries of the system.
We discuss scenarios in which the above
set of fields and couplings appear. The mechanism is quite generic, but a 
natural possibility is that the the $U(1)_X$ symmetry arises from bulk 
gauge bosons in theories with extra dimensions or string models. We show 
that in certain D-brane Type-II string models (with antisymmetric tensor 
field strength fluxes) higher dimensional Chern-Simons couplings give 
rise to the required $D=4$ Wess-Zumino terms upon compactification. In 
one of the possible string realizations of the mechanism the $U(1)_X$ 
gauge boson comes from the Kaluza-Klein reduction of the 
eleven-dimensional metric in M-theory.}

\keywords{Strong CP problem, axions, anomalies, Wess-Zumino terms}

\begin{document}

\section{The strong CP-problem}

The strong-CP problem \cite{cdg,reviews} is one of the oldest fine-tuning 
problems in particle physics.  It is  the statement that the QCD 
${\bar \theta}$-parameter appearing in the action 
\beq
{{\bar \theta}\over {32 \pi^2 }} F_{\mu \nu}^{QCD} { \tilde F}^{\mu \nu}_{QCD}
\label{problem}
\eeq
is indeed  a physically observable  parameter. The presence of such a term 
(which explicitly breaks P and CP) is a consequence of the non-trivial 
structure of the QCD vacuum, and gives rise to computable contributions to 
the electric dipole moment of the neutron which are about ten orders of 
magnitude too large for $\bar \theta $ of order one. Thus one should have 
${\bar \theta} \leq 10^{-10}$. This requires a fine-tuning which gives rise 
to the strong CP problem.

There are a number of proposals to solve the strong CP problem but perhaps 
the most elegant ones are the following two:

\begin{itemize}
\item
{\it A massless quark}.
It is known \cite{thooft,cdg}  that if one of the quarks is massless the 
${\bar \theta }$ phase becomes unobservable, unphysical. This is related 
to the fact that with a massless quark there is a global chiral $U(1)$ 
symmetry preserved by perturbative interactions and violated by the chiral 
anomaly. This is perhaps the simplest solution and it indeed has been 
proposed that the u-quark mass could be zero \cite{km}. However this has 
always been disfavoured by physicists  working on effective chiral 
Lagrangians \cite{leut}. Recent lattice calculations seem also to 
disfavour the possibility of a massless u-quark \cite{lat}.

\item
{\it  The axion solution}

In this solution \cite{pq} the idea is to introduce a dynamical pseudoscalar 
field $\eta^0$ with an axial coupling to the QCD field strength
\beq
{{\eta ^0}\over {f_a }} F_{\mu \nu}^{QCD} { \tilde F}^{\mu \nu}_{QCD}
\label{axionsol}
\eeq
where $f_a$ is a mass parameter which measures the decay width of the 
axion $\eta ^0$. In this mechanism the pseudoscalar $\eta ^0$ (or rather 
$\eta=\eta^0 +{\bar \theta}$) becomes a dynamical `theta parameter'. 
Although the axion is perturbatively massless it acquires a periodic 
scalar potential at the non-perturbative level so that energy is minimized 
at $\eta =0$. Thus the system is relaxed at zero effective 
$\theta$-parameter and  there is no strong CP violation. This is an attractive 
solution but direct searches and astrophysical and cosmological limits 
already rule out most of the parameter space for this model. Only a small 
window with $f_a\propto 10^{10}$ GeV seems to be allowed
\cite{reviews}.

\end{itemize}

\section{Gauging away the strong CP problem}

\subsection{The model}

Our proposal has certain features from both solutions, as will become clear
below. It also borrows some inspiration from string theory.
The key idea is to introduce a $U(1)$ {\em gauge} symmetry, under which 
ordinary quarks are neutral. More especifically, the proposal is to extend 
the SM with

\begin{itemize}

\item {\it A  pseudoscalar} state $\eta$ with axionic couplings to the 
QCD field strength, very much like in the axion solution.

\item{\it A $U(1)_X$ gauge interaction} whose gauge boson gets a 
Stuckelberg mass $M$ by combining with the axion introduced above. This 
means we have a Lagrangian of the form:
\beq
{\cal L} \ =\ {\cal L}_{QCD} \  + \
{\eta} F_{\mu \nu}^{QCD} { \tilde F}^{\mu \nu}_{QCD} \   
- \
\frac{1}{4g^2_X}F^{\mu \nu}_XF_{\mu \nu}^X
- \ {{M^2}\over 2} (A_X^{\mu} \ +\ \partial ^{\mu } \eta)^2
\label{madre}
\eeq
The mass  term is gauge invariant under the transformation
\beq
A_X^{\mu }\rightarrow A_X^{\mu }-\partial ^{\mu } \Theta(x) \ ;\
\eta(x)\rightarrow \eta(x) + \Theta(x)
\label{shiftaxion}
\eeq
\end{itemize}

Instead of a pseudoscalar $\eta$ one can equally consider its Hodge dual, 
a 2-index antisymmetric tensor $B_{\mu \nu }$, representing the same 
degrees of freedom. In this dual language one can write for the relevant 
Lagrangian:
\beq
{\cal L}\ =\  {\cal L}_{QCD} \ -\ \frac{1}{12}H^{\mu \nu \rho}
H_{\mu \nu \rho}\ -\ 
\frac{1}{4g^2_X}F^{\mu \nu}_XF_{\mu \nu}^X\ 
+ \frac{M}{4}  \epsilon^{\mu\nu\rho\sigma} B_{\mu\nu}\ F_{\rho\sigma}^X,
\label{poincar}
\eeq
where 
\beq
H^{\mu \nu \rho}\ =\ \partial^{\mu}B^{\nu\rho}+\partial^{\rho}B^{\mu\nu}
+\partial^{\nu}B^{\rho\mu}
\label{hache}
\eeq
and $F_{\mu\nu}^X$ is the field strength of the $U(1)_X$ gauge field. A 
duality transformation gives back the original Lagrangian in 
eq.(\ref{madre}). 

As it stands this system looks  problematic since the combined presence 
of the $U(1)_X$ transformation of the scalar $\eta(x)$ and the axionic 
coupling implies the presence of a mixed $U(1)_X$-$SU(3)_{QCD}^2$  
anomaly, as depicted in Fig.1-a.

\FIGURE{\epsfxsize=2.6in
\hspace*{0in}\vspace*{.2in}
\epsffile{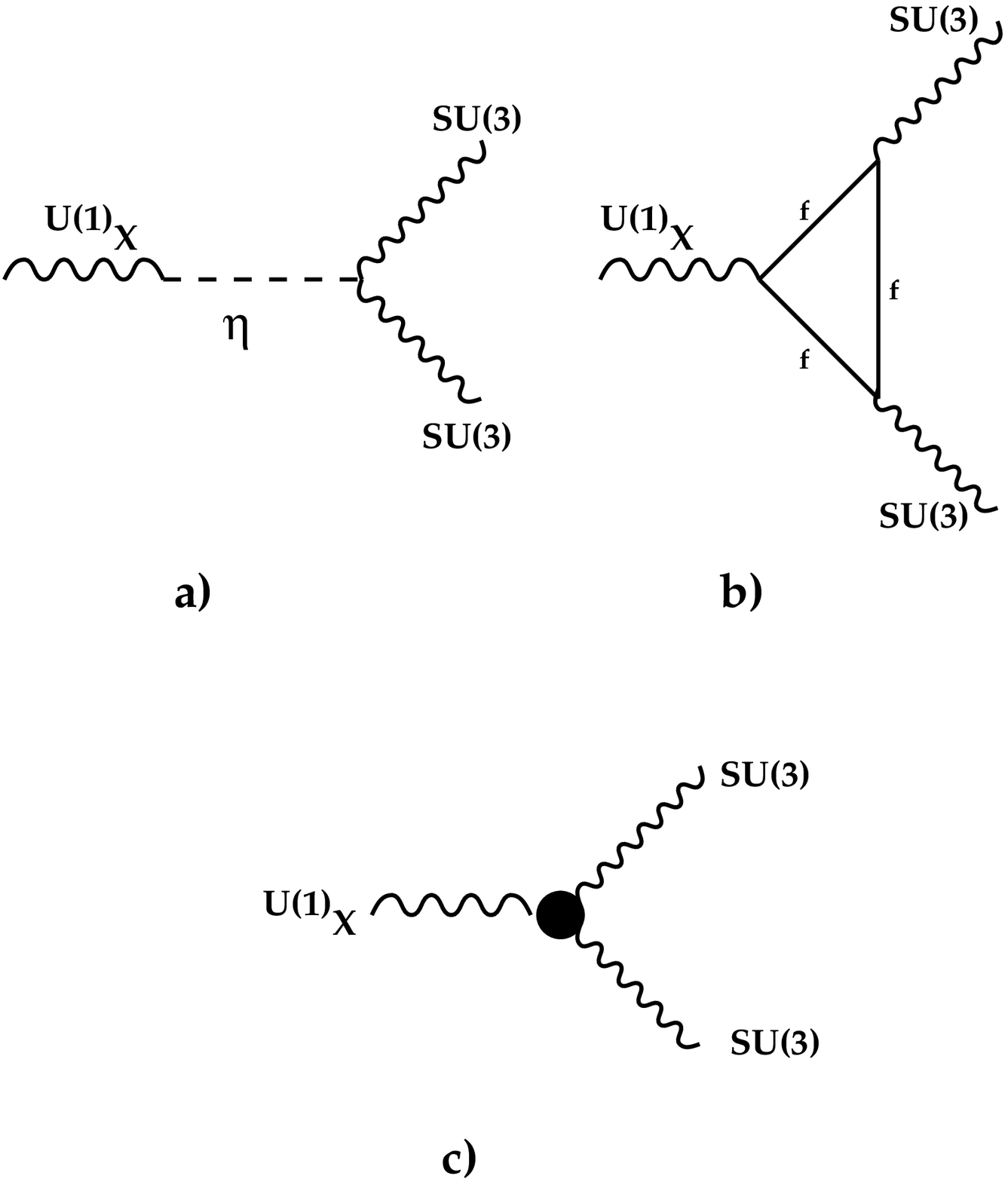}
\caption{\small
Contributions to $U(1)_X\times SU(3)^2$ anomalies:
a) Green-Schwarz contribution from the exchange of the
pseudoscalar $\eta$; b) Standard fermion triangle graph and
c) Contribution from a Wess-Zumino term.}
\label{anomalies}}

An obvious way to cancel this anomaly is to assume the presence of chiral 
fermions which are coloured and charged under the $U(1)_X$. Their 
contribution to the chiral  anomaly (Fig.1-b) may easily cancel the above 
anomalous term . This would be a standard $D=4$ Green-Schwarz mechanism 
in which the axion gauge transformation cancels the mixed 
$U(1)_X$-$SU(3)_{QCD}^2$ anomaly \cite{dsw}. In the case of the SM this would require 
that at least some quark (i.e., the u-quark) is charged and chiral under 
$U(1)_X$. Since we need the $U(1)_X$ symmetry to be unbroken, this means 
that the u-quark will remain massless (zero `current' mass). This we 
would like to avoid since we already mentioned that a massless u-quark is 
disfavoured by  chiral Lagrangian analysis and recent lattice computations.
There are additional reasons to try to avoid the physical quarks being
charged under $U(1)_X$, as we will describe below.

Instead of that we propose that {\it all quarks are neutral} under $U(1)_X$ 
(so that they do not contribute to the mixed anomaly). We also propose that 
the anomaly generated by the axion $\eta(x)$ gauge transformation and the 
axionic coupling is {\it cancelled by a Wess-Zumino term} involving the 
$U(1)_X$ and QCD gauge boson fields (Fig.1-c). Thus we are proposing a 
purely bosonic anomaly cancellation mechanism. Note that the $U(1)_X$ 
gauge boson may be arbitrarily heavy. The only low-enery remnant arises 
from the existence of the Wess-Zumino term, as we discuss later on.

Let us be a bit more concrete about the required Wess-Zumino term. A 
Wess-Zumino term is an explicit non gauge invariant interaction whose 
variation has the structure of a chiral gauge anomaly. Since an anomaly is 
a gauge variation which cannot be cancelled against a local counterterm, 
it is clear that a four-dimensional Wess-Zumino term is non-local 
(although its gauge variation is local) \footnote{In theories with extra 
dimensions, however, non-local four-dimensional Wess-Zumino terms may 
arise from local higher-dimensional interactions, e.g. Chern-Simons terms. 
Hence Wess-Zumino terms of the kind discussed here are more natural in 
higher-dimensional setups, see Section 3.}.

The simplest way to write such terms (see e.g. \cite{manes}) is as follows:
Pick a five-dimensional manifold ${\bf X}_5$ whose boundary is
four-dimensional spacetime $M_4$. Next, extend the four-dimensional gauge 
field to ${\bf X}_5$; that is,
define a five-dimensional gauge field in ${\bf X}_5$ such that it reduces
to the four-dimensional one at the boundary $M_4$. The Wess-Zumino terms
we need are of the form
\beqa
S_{WZ}\, =\, \int_{{\bf X}_5}\, [\, F_{U(1)_X}\, \tr F_{QCD}^2\, ]^{(0)}
\label{wz4d}
\eeqa
Here we are using differential forms (with wedge products implied) and
the Wess-Zumino descent notation. Namely, for a closed gauge-invariant 
anomaly polynomial $Y(F)$, we define $Y=dY^{(0)}$, and $\delta 
Y^{(0)}=dY^{(1)}$, where $\delta$ denotes gauge variation.

The gauge variation of (\ref{wz4d}) gives
\beqa
\delta S_{WZ} \, =\, \int_{{\bf X}_5}\, d\,[\, F_{U(1)_X}\, \tr 
F_{QCD}^2\, 
]^{(1)} \, =\, \int_{M_4}\, [\, F_{U(1)_X}\, \tr F_{QCD}^2\, ]^{(1)}   
\eeqa
which is precisely of the form of a mixed gauge anomaly, and hence cancels 
against the Green-Schwarz contribution mediated by the axion. 

In more pedestrian language, we may write (\ref{wz4d}) as
\beqa
S_{WZ}\, =\, \int_{{\bf X}_5}\, A_X\, \tr F_{QCD}^2
\label{simplewz}
\eeqa
so that its change under a $U(1)_X$ gauge variation $A_X\to A_X + 
d\lambda$ is clearly
\beqa
\delta S_{WZ} \, =\, \int_{{\bf X}_5}\, d\lambda \, \tr F_{QCD}^2\, =
\int_{{\bf X}_5}\, d\, ( \,\lambda \, \tr F_{QCD}^2\,) \, = \, 
\int_{M_4}\, \lambda \, \tr F_{QCD}^2
\eeqa  
This is precisely the contribution required to cancel the gauge variation 
due to the axion shift.

In Section 3 we will present higher-dimensional setups containing 
Wess-Zumino terms in their effective actions. In their discussion the 
Wess-Zumino descent notation turns out to be a bit more convenient, so we 
will stick to it.

\subsection{Gauging away the strong CP problem}

In this section we argue that the symmetries of the above system are
such that the $\theta$ parameter is unphysical.

Since the theta parameter is related to the vev of the axion, and the
latter is shifted by the $U(1)_X$ gauge symmetry, a naive proposal would
be to use the symmetry to shift the axion, and hence its vev, to zero.
However, this idea does not quite work, as is clear from the fact that 
there is no mixed $U(1)$-$SU(3)^3$ anomaly \footnote{We thank D. E. Kaplan 
for comments on this point}. Namely the action must be invariant under 
$U(1)_X$ gauge transformations. In fact, although the $U(1)_X$ 
transformation (\ref{shiftaxion}) shifts the theta parameter, a 
compensating shift arises from the change of the gauge potential in the 
Wess-Zumino term.

The system however has an additional symmetry which we have not
exploited yet, and which does allow to rotate away the theta parameter.
The symmetry is deeply rooted in the structure of the Wess-Zumino term.
In defining it, we need to extend the 4d gauge field to a 5d gauge field on
${X_5}$; namely to define a 5d gauge field on ${\bf X}_5$, four of whose
components reduce to the physical 4d gauge field at the boundary $M_4$.
This still leaves the freedom to choose freely the fifth component. In
particular we are free to choose the constant value of the fifth component
of the gauge field $A_4$ on the boundary. This is an additional $U(1)$
global symmetry of the system, since this component does not appear anywhere
else in the action. As is clear from (\ref{simplewz}), this arbitrary choice
changes the effective value of the theta parameter, showing that it is
indeed unphysical in the system.

More formally, this can be stated as follows. In the quantum theory, one
should path integrate over the 5d gauge field. This implies that, for a
fixed choice of 4d gauge field, we still path integrate over $A_4$ and in
particular over its constant piece at the boundary. This implies that the
quantum theory includes a path integral over the effective 4d theta
parameter, so that its specific value is unphysical, it is not a parameter
of the theory.

The above discussion can be mapped to a perhaps more familiar one by
regarding a Wess-Zumino term as a piece of the (non-local) effective
action arising from integrating out a chiral fermion in a theory.
Specifically, the Wess-Zumino term contains the information concerning the
anomaly properties of such chiral fermion, with respect to gauge and
global anomalies. In our case, the Wess-Zumino term can be regarded as
mimicking a chiral fermion charged under $U(1)_X$ and $SU(3)_{QCD}$.
Indeed the $U(1)$ global phase rotation symmetry of a chiral fermion
corresponds to the global shift of the fifth component of the gauge field
in the Wess-Zumino term. In particular, the anomaly of this global symmetry
in the theory with the chiral fermion is encoded in the explicit change of
the Wess-Zumino terms under a shift of $A_4$. Hence the fact that the theta
parameter of a gauge theory can be removed by a phase rotation of a charged
chiral fermion, corresponds to the statement that the theta parameter can be
removed by a shift of the extra component of the gauge field in the
Wess-Zumino term. In other words, the anomaly of the chiral symmetry of a
massless quarks turns theta into a dynamical variable (the fermion phase),
over which one path integrates in the quantum theory, and whose value is
therefore not a physical parameter of the theory. Analogously, the
Wess-Zumino term turns theta into a dynamical variable (the gauge field
component $A_4$), over which we path integrate in the quantum theory, and
whose value is therefore not a physical parameter of the theory.

Hence, our mechanism to remove the theta parameter is very similar to having
massless fermion, with the important different that we do not have such
a dynamical fermion in the theory, but rather an explicit Wess-Zumino term
which reproduces exactly the same anomaly properties. Notice that the
complete structure of the theory is required for this mechanism to work.
Consider starting with just the Standard Model, and add a Wess-Zumino term to
eliminate the strong CP problem. In order to have a Wess-Zumino term of the
appropriate kind, an additional $U(1)_X$ gauge symmetry is required. This
term then generates mixed gauge anomalies; in order to cancel them without
introducing fermions charged under the $U(1)_X$ symmetry, we need to
implement a Green-Schwarz mechanism, namely introduce an axion coupling to
QCD and mixing with the $U(1)_X$ gauge boson. Hence the model also shares
some features of the axion solution to the strong CP problem. Happily the
mixing of the axion with the $U(1)_X$ gauge boson allows to gauge it away
and avoid inconsistency with experiment.

The key ingredient in the mechanism is the additional $U(1)_X$ gauge sector,
which contains enough symmetries to set to zero both the axion field and
the theta parameter by a combined gauge and global symmetry.
We call this proposal gauging away the strong CP problem.

\subsection{Discussion of other string models}

Axion-like fields transforming under anomalous gauged $U(1)_X$ symmetries
have appeared in the past in $D=4$ string constructions 
\cite{dsw,unos,otros}. However, one of the main differences with our present 
proposal is that, in the specific string models provided in the past, quarks 
and leptons were charged under $U(1)_X$. Thus the fermion contribution to 
gauge anomalies was cancelled by a Green-Schwarz mechanism. In our 
proposal here there are no triangle chiral anomalies: rather, the 
Green-Schwarz contribution cancels against an explicit Wess-Zumino term.

This proposal has a nice advantage. In previously considered string 
models, due to the quarks and leptons being charged, the $U(1)_X$ gauge 
symmetry is always eventually broken in one way or another. For example, 
in the heterotic $D=4$ vacua \cite{dsw,unos} such an axion is ${\rm Im}\, 
S$, the pseudoscalar partner of the dilaton, present in Calabi-Yau or 
orbifold compactifications. In those models there is also a dilaton 
(${\rm Re}\, S$) dependent Fayet-Iliopoulos term associated to the (unique) 
anomalous $U(1)_X$, which forces some scalars charged under it to get a vev 
\cite{unos}. Thus $U(1)_X$ does not survive at low energies in heterotic 
models. A different kind of phenomenological string constructions is 
provided by the explicit D-brane models constructed in the last few years. 
There the situation is in principle slightly better \cite{otros}. There are 
in general more than one anomalous gauged $U(1)$, and it is possible to 
construct D-brane systems in which the $U(1)_X$ remains unbroken (e.g. 
D6-brane intersecting models with positive mass square for all scalars at 
intersections). However this is not sufficient: Since $U(1)_X$ remains an 
unbroken symmetry and typically quarks are charged under it, either some 
quark remains massless (a disfavoured possibility as discussed above) or 
else the Higgs doublets are charged under the residual global $U(1)_X$. 
In the latter case, $U(1)_X$ gets broken in electroweak symmetry breaking, 
spoiling the solution to the CP problem; Moreover, a $U(1)$ with a 
Stuckelberg mass remains as a {\it global} symmetry from the effective 
low-energy theory viewpoint, so its breaking would generate an 
(axion-like) goldstone boson, which is inconsistent with present 
experimental bounds.

The origin of these problems is the fact that in all these string models 
the quarks and leptons were generically charged under the anomalous 
$U(1)_X$'s. To avoid these complications the simplest possibility is to 
assume that {\it  quarks are neutral} under the $U(1)_X$ generator. This 
possibility was not considered before because it was not obvious how to 
cancel the mentioned mixed gauge anomalies. However recently, in the 
context of string compactifications with $p$-form field strength fluxes, 
it has been realized that cancellation of the anomaly may be achieved in 
a purely bosonic manner via a Wess-Zumino term. Those Wess-Zumino terms 
have been shown to appear in explicit D-brane configurations in 
\cite{angelflux}.

One can consider  our proposal to gauge away the strong-CP problem 
independently of any string theory or extra dimension argument. However  
natural candidates for $U(1)_X$ bosons, in brane-world models with extra
dimensions, are bulk gauge fields, which have
no couplings to brane chiral fermions. In this way the usual quarks and 
leptons (which live on the D-braves) are neutral under  $U(1)_X$. In what 
follows we will describe how this structure may naturally appear in models 
with extra dimensions. In particular we will show how the required 
couplings and fields appear in explicit D-brane constructions.

\section{Some examples from extra dimensions and string theory}

\subsection{ Wess-Zumino terms from higher dimensions}

It is easy to understand that Wess-Zumino terms of the kind needed above 
can easily appear in theories with extra dimensions. The reason for this 
is that non-local four-dimensional Wess-Zumino terms may arise from local 
operators in higher dimensions. For instance, five-dimensional Chern-Simons 
terms roughly of the form $\int_{X_5} [ \tr F^3]^{(0)}$ have appeared in 
five-dimensional orbifold models \cite{orbifield} in order to cancel 
the four-dimensional anomaly generated by chiral fermions at the fixed 
points of the orbifold (i.e. boundaries of the five-dimensional space 
${\bf X}_5$). From the perspective of the four-dimensional boundary
such interactions behave as $D=4$ Wess-Zumino terms.

Here we would like to show that higher dimensional Chern-Simons interactions 
(of a different kind) also lead to four-dimensional Wess-Zumino terms, 
in general compactifications of field theories with $p$-form field 
strength fluxes. Consider a $(p+4)$-dimensional theory with the QCD and 
$U(1)_X$ gauge bosons propagating in the bulk. Consider the space is 
compactified to four dimensions on a $p$-dimensional manifold ${\bf 
X}_p$. Also introduce a $(p-1)$-index antisymmetric tensor field 
$C_{p-1}$, whose field strength $H_p$ has non-zero (quantized) flux over 
${\bf X}_p$, 
\beqa
\int_{X_p} H_p = k_{\rm flux} \; \in \IZ
\eeqa
and which interacts with the $U(1)_X$ and $SU(3)_c$ gauge bosons via a 
Chern-Simons coupling
\beqa
S_{CS}\, = \, \int_{M_4\times X_p}\, C_{p-1}\, A_X\, \tr F_{QCD}^2
\label{chernsimons}
\eeqa
Here we are using differential form notation, with wedge products implied.
 Using Wess-Zumino descent notation, this may be written as 
\beqa
S_{CS}\, = \, \int_{M_4\times X_p}\, C_{p-1}\, [\,F_X\, \tr F_{QCD}^2\,]^{(0)}
\eeqa
This term is not invariant under $U(1)_X$, $SU(3)_c$ gauge 
transformations, its variation being given by
\beqa
\delta S_{CS} \, \simeq \,  \int_{M_4\times X_p}\, H_{p}\, [\,F_X\, \tr 
F_{QCD}^2\,]^{(1)}\, =\, k_{\rm flux}\, \int_{M_4} [\,F_X\, \tr
F_{QCD}^2\,]^{(1)}
\eeqa
Hence it behaves exactly as a four-dimensional Wess-Zumino term of the 
required kind. That is, it generates a four-dimensional gauge variation of 
precisely the form required to cancel the mixed $U(1)_X$-$SU(3)_c^2$ 
anomaly generated  by the Green-Schwarz contribution.

The above ingredients, $p$-form field, interactions, etc, have been 
introduced in a rather {\em ad hoc} fashion. In the following we discuss 
that these ingredients, and this mechanism, are automatically present in 
large classes of string compactifications with D-branes and fluxes.

\bigskip

\subsection{Wess-Zumino terms in string theory}

In this section we show that Wess-Zumino terms of the kind discussed above 
are naturally present in large classes of type II string compactifications 
with $p$-form field strength fluxes. Such compactifications have recently 
received attention \cite{fluxes} since they lead to other phenomenologically 
interesting properties, for instance they provide mechanisms of moduli 
stabilization. 

On the other hand, we are interested in models with phenomenologically 
appealing spectra, hence including non-abelian gauge symmetries and 
chiral fermions. In order to achieve this, we consider compactifications 
with D-branes; specifically we consider type IIA compactification on a 
six-dimensional manifold ${\bf Y}_6$, with D6-branes wrapped on 3-cycles on 
${\bf Y}_6$. Standard model gauge interactions propagate on the volumes of 
the different D6-brane stacks, while four-dimensional chiral fermions 
arise at intersections of the D6-branes \cite{bdl}. Phenomenological 
compactifications of this kind with ${\bf Y}_6$ a six-torus (or 
orbifold/orientifold quotients thereof) have been studied in \cite{orbif, 
bgkl, afiru,rest,susy} \footnote{See \cite{phenodbrane1, 
phenodbrane2} for other phenomenological D-brane model building setups. 
In principle it should be possible (and interesting) to implement the 
gauging away of the CP problem in these alternative setups.}.

Hence we consider type IIA theory compactified on ${\bf Y}_6$, with $K$ 
stacks of $N_a$ coincident  D6-branes, $a=1,\ldots,K$, wrapped on 3-cycles 
$[\Pi_a]$ in ${\bf Y}_6$ \footnote{An explicit realization of these brane 
configurations in which ${\bf Y}_6$ is a 6-torus is given in the 
appendix.}. Quantization of the open string sectors leads to $U(N_a)$ 
gauge interactions propagating on the volume of the D6$_a$-branes, and 
chiral four-dimensional fermions in the bi-fundamental representation 
$(N_a,{\ov N}_b)$ at the intersections of the 3-cycles $[\Pi_a]$, 
$[\Pi_b]$ in ${\bf Y}_6$. Such fermions arise with multiplicity given by 
the number of intersections $I_{ab}=[\Pi_a]\cdot [\Pi_b]$. The closed 
string sector contains several $p$-index antisymmetric tensor fields, the 
RR $p$-forms, $C_1$, $C_3$, $C_5$, $C_7$, which can lead to 
four-dimensional 1-forms upon compactification. These fields may easily 
play the role of the $U(1)_X$ gauge boson $A_X$ in our above mechanism. 
In the following we describe this in the case of the type IIA RR 1-form 
$A_X=C_1$.

First we need to identify the QCD axion in our setup. In Type IIA string 
theory the gauge fields on the D6-brane couple to the closed string RR 
modes via Chern-Simons couplings. Among them we have 
\beqa
\int_{D6_a}\, C_5\, F_a \quad ; \quad \int_{D6_a} C_3 \, \tr F_a^{\, 2}
\label{branosas}
\eeqa
where products in all equations are exterior products.
From the four-dimensional perspective, defining 
\footnote{In general the 
fields $\eta_a$ are not linearly independent. In order to work with 
independent fields, we may choose a basis of 3-cycles $[\Sigma_i]$, 
decompose $[\Pi_a]=\sum_i n_{ai} [\Sigma_i]$, and define 
$\eta_i=\int_{\Sigma_i} C_3$. The interaction then reads $\sum_i 
n_{ia}\int \eta_i \tr F_a^2$. We skip the subtlety for clarity.} 
$\eta_a=\int_{\Pi_a} C_3$, the second interaction becomes
\beqa
\int_{M_4} \eta_a \, \tr F_a^2
\label{axioncete}
\eeqa
So if QCD arises from a stack of D6-branes wrapped on a given cycle 
$[\Pi_{a_0}]$, then $\eta_{a_0}$ is the QCD axion. For future use we 
notice that the $\eta _a$ degree of freedom may be also represented by its 
four-dimensional Hodge dual, given by  $B_2^a=\int_{\tilde{\Pi}_a} C_5$, 
where $[\tilde{\Pi}_a]$ is the cycle dual to $[\Pi_a]$. 

On the other hand, in compactifications with non-zero flux for the NS-NS 
3-form field strength $H_{NS}$, the axions $\eta_a$ have  Stuckelberg 
couplings with bulk gauge fields via the four-dimensional reduction of the 
type IIA ten-dimensional interaction
\beqa
\int_{M_4\times Y_6} \, C_5 \, H_{NS} F_2
\label{bulkmix}
\eeqa
where $F_2$ is the field strength of the type IIA RR 1-form, $C_1$. 
Now, if we  turn on a flux of $H_{NS}$ along $[\Pi_{a_0}]$, 
$\int_{\Pi_{a_0}} H_{NS}=k_{\rm flux}$, the term in  (\ref{bulkmix}) 
gives rise to a coupling 
\footnote{More generally, the duals of $\eta_i$ are 
$B_2^j=\int_{\Lambda_j} C_5$ with $\Lambda_j$ are a basis of 3-cycles dual 
to $\Sigma_i$, namely $[\Sigma_i]\cdot[\Lambda_j]=\delta_{ij}$. In 
compactifications with general fluxes $\int_{\Sigma_i} H_{NS}=k_i$ there 
are, among others, induced couplings $k_i \int_{M_4} B_2^i F_2$.}
\beqa
k_{\rm flux} \int_{M_4} B_2^{a_0} F_2
\label{bfbulk}
\eeqa
Note that this coupling is analogous to the last term in (\ref{poincar}).
Thus, the bulk RR 1-form field plays the role of $U(1)_X$, and its gauge 
invariance makes the theta parameter for the QCD $U(N_{a_0})$ group 
unobservable. 

Note that in general the antisymmetric form $B_2^{a_0}$ may also have 
similar $B_2^{a_0}\wedge F_2^{brane}$ couplings with the gauge bosons 
living on the D6-branes. Those appear after dimensional reduction from 
the first equation in (\ref{branosas}), taking into account that
$B_2^{a_0}=\int_{\tilde{\Pi}_{a_0}} C_5$. 
These couplings are dangerous if we want the mechanism to solve the strong 
CP problem to work. The reason is that if both those couplings and the 
ones in (\ref{bfbulk}) are present, the bulk $U(1)$ field will mix with 
the $U(1)$'s on the branes. Since the brane fermions (like e.g. quarks in
a realistic model) are charged with respect to the brane $U(1)$'s, they 
will also acquire charges with respect to the bulk $U(1)$
(more correctly, the different BF couplings induce mixing among the 
different $U(1)$'s, so that in general $U(1)_X$ is a mixture of bulk and 
brane $U(1)$'s, under which the quarks are in general charged). As we 
discussed in previous sections, this is something we would like to avoid. 
As we will show in the specific example in the appendix, it is easy to 
find D6-brane configurations in which this mixing of bulk and brane 
$U(1)$'s is absent.

As discussed, the combination of the couplings (\ref{axioncete}) and 
(\ref{bfbulk}) gives rise to mixed $U(1)_X$-$U(N_a)^2$ anomalies. However 
we now show that they are cancelled by Wess-Zumino terms which are present 
in the theory. Indeed, the model contains Chern-Simons interactions which 
generate the adequate four-dimensional WZ terms. In fact, following 
\cite{angelflux} the interaction
\beqa
\int_{D6} (C_1+C_3+...)\; e^{F+B_{NS}} 
\eeqa
on the D6-brane world-volume leads to a coupling 
\beqa
S_{CS}=\int_{D6_{a_0}}\, B_{NS}\, C_1\, \tr F_{a_0}^2 \, = 
\int_{D6_{a_0}}\, B_{NS}\, [\,F_2\, \tr F_{a_0}^2 \, ]^{(0)}
\eeqa
where $B_{NS}$ is the NS-NS 2-index antisymmetric tensor field. This is 
precisely of the form (\ref{chernsimons}). Hence, its gauge variation is
\beqa
\delta S_{CS} = \int_{D6_{a_0}}\, H_{NS} \, [\,F_2\, \tr F_{a_0}^2 \, 
]^{(1)} \, 
= \, k_{\rm flux} \int_{M_4}  [\,F_2\, \tr F_{a_0}^2 \, ]^{(1)}   
\eeqa
and provides the required term  to cancel Green-Schwarz contribution from  
 eq.(\ref{axioncete}) and (\ref{bfbulk}).

In the appendix we present concrete examples of this mechanism 
in explicit string constructions with standard model like spectrum.

It is possible to implement the above mechanism using four-dimensional 
bulk gauge modes arising from compactification of higher degree $p$-forms 
in IIA theory. It is also easy to describe the mechanism in 
type IIB compactifications with fluxes and D-branes. 

An amusing feature of the particular realization we have described is the 
following. Note that the gauge boson $U(1)_X$ comes the Type-IIA RR 
1-form, $C_1$. If we do the lift to M-theory such 1-form comes from the 
circle compactification of the mixed component of the eleven-dimensional 
metric, i.e. $C_{\mu}=g_{\mu 11}$. The $U(1)_X$ gauge invariance arises 
from local translation invariance of the circle of the 11-th dimension. 
Hence in the above setup the $\theta$-parameter has been gauged away using 
diffeomorphism invariance in M-theory.

\section{Final comments}

We have proposed a new solution to the strong-CP problem. The mechanism 
involves the gauging of a $U(1)_X$ symmetry whose boson gets a Stuckelberg 
mass by combining with an axion-like field $\eta(x)$. The latter has 
axionic couplings to $F_{QCD}\wedge F_{QCD}$.
The mass of the combined system axion-gauge boson is arbitrary.
The  combined system leads to mixed gauged anomalies 
$U(1)_X$-$SU(3)_{QCD}^2$, cancelled by Wess-Zumino terms which
should be provided by the underlying theory.
The axion can be gauge away using the $U(1)_X$ gauge transformation,
while the QCD theta parameter can be removed by a shift of the extra
component of the gauge field in the Wess-Zumino term. Hence the strong
CP problem is solved by the specific symmetries of the additional $U(1)_X$
gauge sector.

The required extra fields, a $U(1)_X$ anomalous gauge boson and a 
pseudoscalar which provides its longitudinal degrees of freedom, naturally 
appear in models with extra dimensions. This is also the case of the 
required Wess-Zumino terms cancelling anomalies. We have shown how certain 
higher dimensional Chern-Simons couplings provide us with the apropriate 
Wess-Zumino terms in the presence of certain antisymmetric tensor field 
fluxes. As an example we have shown that simple type II string 
compactifications in the presence of D-branes have all the ingredients for 
the mechanism to work if certain antisymmetric field fluxes are present. 
An explicit Type II toroidal example with a configuration of intersecting 
D6-branes yielding a semirealistic three generation model is provided in 
which the strong-CP problem is gauged away. It is interesting that in the 
string examples which we have discussed the $U(1)_X$ symmetry corresponds 
to the RR 1-form of Type IIA string theory. Thus this $U(1)_X$ gauge 
symmetry admits a geometrical interpretation as part of the 
reparametrization invariance of eleven-dimensional  M-theory.

\bigskip

\bigskip

\bigskip

\centerline{\bf Acknowledgements}

We thank D. Cremades, C. Fosco, F. Marchesano, J.~F.~Garc\'{\i}a Cascales,
F. Quevedo   and R. Rabad\'an for useful 
discussions. A.M.U. thanks M.~Gonz\'alez for kind encouragement and 
support. This work has been partially supported by CICYT (Spain), the 
European Commission (grant ERBFMRX-CT96-0045), G.A work is partially 
supported by ANPCyT grant 03-03403. A.M.U. is supported by the Ministerio 
de Ciencia y Tecnolog\'{\i}a (Spain) via a Ram\'on y Cajal contract.
G.A. thanks U.A.M for hospitality.

\newpage

\appendix

\section{Appendix I: An explicit D-brane example }

In this appendix we show how the above ideas can be implemented in a Type 
IIA string theory context with D6-branes at angles. We start by briefly 
summarizing some facts about such D6-brane models 
(see ref.\cite{bgkl,afiru,rest} for details).  

We consider type IIA theory compactified on a factorizable six-torus
$\IT^6=(\IT^2)_1 \times (\IT^2)_2 \times (\IT^2)_3$, product of three 
two-dimensional tori. Each such two-torus $(\IT^2)_I$ ($I=1,2,3$), taken 
rectangular for simplicity, is obtained as a quotient of $\IR^2$ by 
lattice translations generated by unit vectors ${\bf e_1}^I=(1,0)^I$ and 
${\bf e_2}^I=(0,1)^I$. 

We also introduce $K$ sets of $N_a$ ($a=1\dots,K$) coincident D6-branes 
wrapped on 3-cycles of $\IT^6$, constructed as a factorized product of 
three one-cycles on each of the three two-tori $(\IT^2)_I$. Thus, each set 
of branes defines the wrapping numbers $(n_a^I, m_a^I)$ on each 
$(\IT^2)_I$, $I=1,2,3$, namely it spans a one-cycle in $(\IT^2)_I$ 
wrapping $n_a^I$ times around the ${\bf e_1}^I$ direction and $m_a^I$ 
times 
around the ${\bf e_2}^I$ direction. Therefore, the angle of these branes 
with 
the ${\bf e_1}^I$ axis is given by
\begin{equation}\label{angle}
\tan\vartheta _a^I= \frac{m_a^I R_2^I}{n_a^I R_1^I}
\end{equation}
where $ R_1^I,R_2^I$ are the two-tori radii. Such considerations are 
easily generalized to  skewed two-tori.

Open strings stretching within the same set of $N_a$ D$6_a$-branes give 
rise to a $U(N_a)$ gauge group \footnote{Actually, if the wrapping numbers 
on a given two-torus $(n,m)$ are not coprime,  the world-volume gauge group 
is $U(N_a/r)^{r}$ with $r=\gcd(n,m)$ the greatest common divisor 
\cite{afiru}.}. The chiral spectrum comes from  strings stretching between 
branes in different sets. Thus, the gauge group and chiral fermion 
spectrum read 
\beqa
&  \prod_{a=1}^K U(N_a)& \nonumber \\
& \sum_{a<b}\, I_{ab}\, (N_a,{\ov N}_b)&
\label{spec6}
\eeqa
where $I_{ab}$ 
\beqa
I_{ab}=[\Pi_a]\cdot [\Pi_b] =\prod_i(n_a^i m_b^i-m_a^i n_b^i)
\label{defintersec}
\eeqa
counts the number of intersections.

Cancellation of RR tadpoles $\sum_a N_a [\Pi_a]=0$ requires the wrapping 
numbers to satisfy the constraints
\beqa
\begin{array}{lcl}
\sum_a N_a n_a^1 n_a^2 n_a^3 = 0 & \quad &
\sum_a N_a n_a^1 m_a^2 m_a^3 = 0 \\
\sum_a N_a m_a^1 n_a^2 n_a^3 = 0 & \quad &
\sum_a N_a m_a^1 n_a^2 m_a^3 = 0 \\
\sum_a N_a n_a^1 m_a^2 n_a^3 = 0 & \quad &
\sum_a N_a m_a^1 m_a^2 n_a^3 = 0 \\
\sum_a N_a n_a^1 n_a^2 m_a^3 = 0 & \quad &
\sum_a N_a m_a^1 m_a^2 m_a^3 = 0
\end{array}
\label{tadpole}
\eeqa
which ensure  the cancellation of cubic non-abelian anomalies, which  
for the $SU(N_a)$ factor in (\ref{spec6}) read 
\beqa
\sum_{b=1}^K\, I_{ab}\, N_b = 0
\label{anomdsix}
\eeqa

Gauge fields from D6-branes, wrapped on 3-cycles, couple to RR (pseudo)-
scalar $\eta _i$ fields and to their 2-form duals $B_i^{(2)}$ as discussed 
in section 3.2. Let us be more explicit for the $\IT^6$ case we are 
considering, and classify the different axion like fields by indicating 
which basis 3-cycle they arise from (see footnote 4). 
Namely,  
\beqa
\begin{array}{cclccl}
\eta _{123} & = & \int_{{\bf e_1}^ 1 \otimes {\bf e_1}^ 2\otimes {\bf 
e_1}^ 3}
C_3\\
\eta_I  & = & \int_{{\bf e_1}^ I \otimes {\bf e_2}^ J\otimes {\bf e_2}^ K}
C_3\\
\eta_{IJ}  & = & \int_{{\bf e_1}^ I \otimes {\bf e_1}^ J\otimes {\bf e_2}^ 
K}
C_3\\
\eta   & = & \int_{{\bf e_2}^ 1 \otimes {\bf e_2}^ 2\otimes {\bf e_2}^ 3}
C_3
\end{array}
\eeqa
where $I\neq J\neq K\neq I$ in the second and third rows.

The Hodge dual 2-forms are defined accordingly. For instance, the Hodge 
dual for $\eta _{123}$ is 
\beq
B_{123}^{(2)} = \int_{{\bf e_2}^ 1 \otimes {\bf e_2}^ 2\otimes {\bf e_2}^ 
3}
C_5
\eeq
Thus, for D$6_a$-branes with wrappings numbers $(n_a^I, m_a^I)$ the 
following couplings between RR fields and brane gauge bosons are 
obtained
\beqa
\begin{array}{ccc}
\label{RRD6c}
N_a\, m^1_a\, m^2_a\, m^3_a \int_{M_4} B_{123}^{(2)} \wedge F_a & \quad ; 
\quad
& n^1_b\, n^2_b\, n^3_b \int_{M_4} \eta_{123} \wedge F_b \wedge F_b \nonumber \\
N_a\, n^J_a\, n^K_a\, m^I_a \int_{M_4} B_{I}^{(2)} \wedge F_a & \quad ; 
\quad
& n^I_b\, m^J_b\, m^K_b \int_{M_4} \eta_{I} \wedge F_b\wedge F_b  \nonumber\\
N_a\, n^K_a\, m^I_a\, m^J_a \int_{M_4} B_{IJ}^{(2)} \wedge F_a & \quad ; 
\quad
& n^I_b\, n^J_b\, m^K_b \int_{M_4} \eta_{IJ} \wedge F_b\wedge F_b \nonumber \\

N_a\, n^1_a\, n^2_a\, n^3_a \int_{M_4} B^{(2)} \wedge F_a & \quad ; \quad
& m^1_b\, m^2_b\, m^3_b \int_{M_4} \eta \wedge F_b\wedge F_b
\end{array}
\eeqa

In order to achieve the mechanism for gauging away the $\theta$-parameter 
we must identify an axion-like field $\eta$ which couples to QCD through 
a term $\eta\, \tr F_{QCD}^2$, and whose Hodge dual has BF couplings to 
a bulk RR 1-form field $U(1)_X$ as in section 3.2. As stressed above,  
couplings of the dual axion field with $U(1)$ gauge bosons from  
D$6_a$-branes must be avoided.

In what follows a Standard like model with these properties, obtained from  
D6-branes at angles, is presented. It contains six sets of D6-branes with 
$N_1=3$, leading to QCD group, $N_2=2$ and $N_3= N_4=N_5=N_6=1$. Wrapping 
numbers are given in Table~\ref{tabps4} and lead to the following, non-zero, 
intersection number
\bigskip
\bigskip

\TABLE{
\renewcommand{\arraystretch}{0.99}
\begin{tabular}{|c||c|c|c|c|}
\hline
$N_1=3$&  (1,0) &  (1,-1) &   (1,-1) \\ 
\hline
$N_2=2$ &   (2,1)&   (1,2)&    (1,0)\\ 
\hline
$N_3=1$ & (2,1)&   (-1,-2)&   (1,0)\\
\hline
$N_4=1$&  (1,0) &  (1,-1) &   (-2,2) \\ 
\hline
$N_5=1$ &   (1,0)&   (1,-1)&    (-1,1)\\ 
\hline
$N_6=1$ & (2,1)&   (-1,-2)&   (1,0)\\
\hline \end{tabular} 
\caption{ Wrapping numbers}
\label{tabps4}}
\bigskip
\bigskip

\beqa
I_{12} &=& 3=I_{56}=I_{25} \\ \nonumber
I_{13} & = &I_{16}=I_{35} =-3\\ \nonumber
I_{24}&=&I_{46}=6= -I_{34} \nonumber
\eeqa

\bigskip
\bigskip
\bigskip

By recalling eq.(\ref{spec6})  we find the gauge group 

\beq 
SU(3) \times SU(2)\times U(1)_Y  (\times U(1)'s)
\eeq
with the chiral fermion spectrum
\beqa \nonumber
&& 3(3,2,1/6)_{(1,-1,0,0,0,0)}+ 3({ \bar 3},1,-2/3)_{(-1,0,1,0,0,0)}+  
3({ \bar 3},1,1/3)_{(-1,0,0,0,0,1)}+\\ \nonumber
&&3(1,2,1/2)_{(0,1,0,0,-1,0)}+ 3(1,2,-1/2)_{(0,1,0,{\underline {-1,0}},0,0)}+\\ \nonumber
&&3(1,1,1)_{(0,0,0,{\underline {1,0}},0,-1)}+3(1,1,-1)_{(0,0,-1,0,1,0)}+\\ \nonumber
&& 3(1,1,0)_{(0,0,0,0,1,-1)}+3(1,1,0)_{(0,0,-1,{\underline {1,0}},0,0)}
\eeqa 
where undelining means permutation (notice multiwrapping on third torus 
for $N_4$ ).
Hypercharge is defined as a linear combination of $U(1)$ generators
\begin{eqnarray}
\label{hypercharge}    
Y &=&  -(\frac{Q_1}3 +\frac{Q_2}2+Q_5+Q_6)
\end{eqnarray}
where $Q_a$ is the $U(1)$ generator in $U(N_a)$.

From the wrapping numbers presented in Table~\ref{tabps4}, it follows that 
$\eta_{123}$ has an axion coupling
\footnote{The careful reader may notice that the QCD axion field is 
actually a linear combination of $\eta_{123}$ and other RR fields; 
however, this is enough to gauge away the QCD $\theta$-parameter, since 
the gauge $U(1)_X$ symmetry used below shifts the total axion via its 
shift on $\eta_{123}$.}
to QCD through the first term in (\ref{RRD6c})
\beq
\int_{M_4} \eta_{123} \wedge \tr\, (F_{QCD} \wedge F_{QCD}) \nonumber 
\eeq
Moreover, the dual 2-form $B_{123}^{(2)}$ has no BF coupling to brane 
gauge fields, since $ m^1_a\, m^2_a\, m^3_a =0$ for all $a$, ($a=1,\dots,6$). 

Finally, by turning on $k_{\rm flux}$ units of flux 
\footnote{The conditions on allowed field-strength $p$-form fluxes have 
been well studied in the literature \cite{fluxes}. In general it is 
possible to complement the above choice of NS-NS flux with additional 
RR fluxes so that the complete set of fluxes satisfies the closed string 
equations of motion (at least for certain choices of geometric/dilaton 
moduli). We skip this discussion since it is irrelevant for our main 
point.}
for $H_{NS}$ along the directions 
${\bf e_1}^1$, ${\bf e_1}^2$, ${\bf e_1}^3$ the dual 2-form has the 
required BF coupling with the bulk type IIA RR 1-form $C_1$, via 
(\ref{bulkmix})
\beqa
\int_{M_4\times T^6} C_5 F_2 H_{NS} = k_{\rm flux} \int_{M_4} 
B^{(2)}_{123} F_2
\eeqa

Therefore the above model is an explicit stringy example where the 
gauging away $\theta$ mechanism proposed in the article is realized.

This is quite remarkable since the model we have discussed is indeed 
really close to the structure of the Standard Model. On one hand the 
spectrum contains the minimal Standard Model fermions, and a not too large 
set of extra fields, namely six extra doublets and some zero hypercharge 
singlets. This extra matter is expected from tadpole cancellation (see 
\cite{imr} for a discussion). Extra doublets and at least part of these 
singlets may become massive through a Higgs mechanism, and can be absent 
in other constructions. The gauge group is also close to the Standard 
Model, plus some $U(1)$'s. However, several of the $U(1)$'s become 
massive due to the BF couplings (\ref{RRD6c}) \cite{afiru,imr}. This 
makes some of the D6-brane $U(1)$'s massive, so they disappear from the 
low-energy physics. Fortunately, the linear combination 
(\ref{hypercharge}) which we identified with hypercharge in the above 
model can be checked to be free of BF couplings, hence it remains massless 
and part of the Standard Model gauge group.

\end{document}